\documentclass[twocolumn,english,showpacs,superscriptaddress,prl,preprintnumbers,amsmath,amssymb,floatfix]{revtex4-1}
\usepackage[T1]{fontenc}
\usepackage[utf8]{inputenc}
\usepackage{babel}
\usepackage{amsmath}
\usepackage{graphicx}
\usepackage{esint}
\usepackage[unicode=true,pdfusetitle,
 bookmarks=true,bookmarksnumbered=false,bookmarksopen=false,
 breaklinks=false,pdfborder={0 0 1},backref=false,colorlinks=false]
 {hyperref}

\makeatletter

\usepackage{dcolumn}
\usepackage{bm}
\usepackage{color}
\usepackage{soul}
\usepackage{babel}
\usepackage{amsfonts}
\usepackage{slashed}
\usepackage{enumerate}
\graphicspath{{./}}

\usepackage{pdfpages}
\AtBeginDocument{\let\LS@rot\@undefined} 

\makeatother

\usepackage{babel}
\begin{document}
\global\long\def\sgn{\mathrm{sgn}}

\title{Measurement and control of a Coulomb-blockaded parafermion box}

\author{Kyrylo Snizhko}

\affiliation{Department of Condensed Matter Physics, Weizmann Institute of Science,
Rehovot, 76100 Israel}

\author{Reinhold Egger}

\affiliation{Institut f{ü}r Theoretische Physik, Heinrich-Heine-Universit{ä}t,
D-40225 D{ü}sseldorf, Germany}

\author{Yuval Gefen}

\affiliation{Department of Condensed Matter Physics, Weizmann Institute of Science,
Rehovot, 76100 Israel}

\date{\today}
\begin{abstract}
Parafermionic zero modes are fractional topologically protected quasiparticles
expected to arise in various platforms. We show that Coulomb charging
effects define a parafermion box with unique access options via fractional
edge states and/or quantum antidots. Basic protocols for the detection,
manipulation, and control of parafermionic quantum states are formulated.
With those tools, one may directly observe the dimension of the zero-mode
Hilbert space, prove the degeneracy of this space, and perform on-demand
digital operations satisfying a parafermionic algebra.
\end{abstract}
\maketitle
\emph{Introduction.} Majorana zero modes are canonical examples for
topologically protected quasiparticles with non-Abelian braiding statistics
\cite{Nayak2008,Alicea2012,Leijnse2012,Beenakker2013}. In the presence
of Coulomb charging effects, intriguing features related to their
nonlocality have been pointed out \cite{Fu2010,Beri2012,Altland2013,Beri2013,Landau2016,Aasen2016,Vijay2016,Plugge2017,Karzig2016}
and probed experimentally \cite{Albrecht2016,Albrecht2017}. The drive
for reaching universal quantum computation platforms and the quest
to fully understand topological excitations have turned attention
to exotic emergent quasiparticles such as parafermions (PFs) with
$\mathbb{Z}_{n>2}$ symmetry. For PF zero modes, a plethora of interesting
phenomena has been suggested in various platforms \cite{Lindner2012,Cheng2012,Clarke2013,Burrello2013,Vaezi2013,Zhang2014,Mong2014,Clarke2014,Barkeshli2014a,Barkeshli2014b,Cheng2015,Alicea2015a,Kim2017};
for a review, see Ref.~\cite{Alicea2016}. For one-dimensional (1D)
interacting fermions, Ref.~\cite{Fidkowski2011} suggests that Majorana
states exhaust all possibilities in the generic (disordered) case.
However, Refs.~\cite{Klinovaja2014,Klinovaja2014b} show that PFs
can exist in models of 1D nanowires. Moreover, for edge states of
a fractionalized two-dimensional system, such as the fractional quantum
Hall (FQH) liquid, domain walls between regions proximitized by a
superconductor (SC) \cite{foot:SC+magnetic_field} and a ferromagnet
(FM) host stable PFs. Platforms for PFs include proximitized fractional
topological insulators \cite{Lindner2012}, bilayer FQH systems \cite{Barkeshli2014b},
and proximitized FQH liquids at a filling factor of $\nu=2/3$ \cite{Mong2014}
or $\nu=1/(2k+1)$ with an integer $k$ \cite{Lindner2012,Clarke2013}.
Such setups may ultimately provide a toolbox for generating Fibonacci
anyons \cite{Mong2014,Alicea2015a}, which, in turn, facilitate fault-tolerant
universal quantum computation.

\begin{figure}[t]
\centering \includegraphics[width=1\columnwidth]{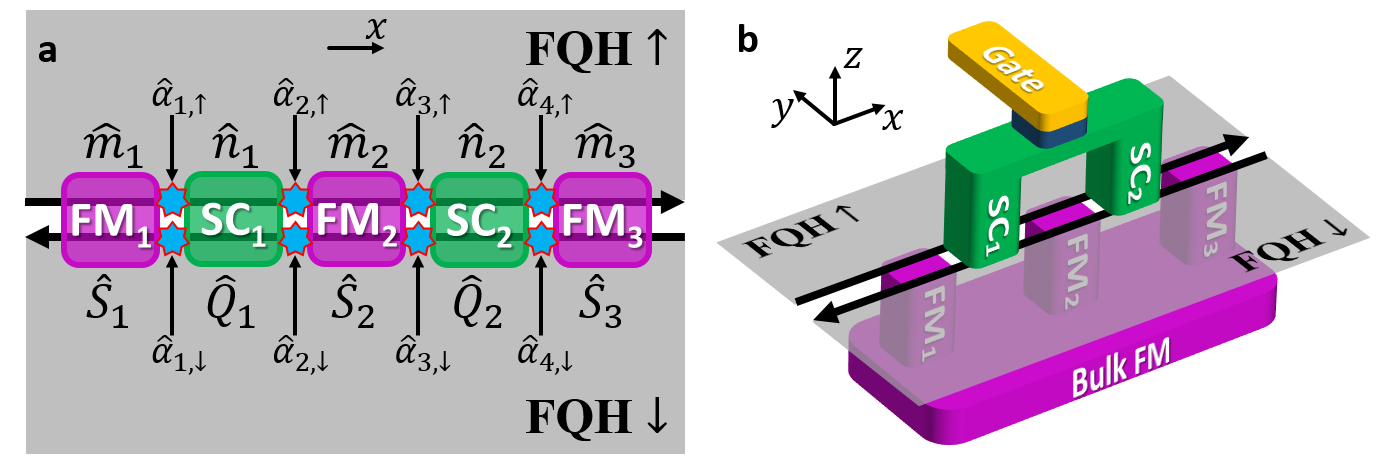} \caption{\label{fig1} Layout of the PF box. (a) Two opposite-spin FQH edges
(the thick black arrows) are gapped out in distinct ways via proximitizing
FM and SC segments in different regions. PF operators $\hat{\alpha}_{j,s}$
(the blue stars) are localized at domain walls. (b) SC segments are
electrically connected to ensure phase coherence across different
SC domains. Similarly, FM segments belong to one bulk FM. The gate
(golden), separated (insulating blue layer) from the SC bridge, can
be used to change the offset charge $q_{\mathrm{g}}$ while disturbing
the rest of the system only minimally. A back gate that can tune the
chemical potential of the FQH puddles is implied.}
\end{figure}

In the present Rapid Communication we leap beyond the interesting
platforms alluded to above. We point out that PF devices dominated
by Coulomb charging effects provide direct detection and manipulation
tools targeting the fundamental physics of PFs. Specifically, we show
below how one can: (i) measure the dimension of the Hilbert space
associated with PF zero modes, (ii) render this space degenerate in
a controlled manner, and (iii) explicitly demonstrate the exotic algebra
of PF operators. By combining systems made of fractionalized bulk
matter with mesoscopic sensing concepts, all superimposed on Coulomb
charging effects, the PF box (cf. Fig.~\ref{fig1}) facilitates full
access to the beautiful physics of PF zero modes. Recent Majorana
experiments \cite{Albrecht2016,Albrecht2017} also attest to the promise
of such an approach. Probing these core facets of PF Hilbert space
is realized employing fractional edge states {[}for current measurements{]}
and quantum antidots (QADs) {[}for elastic cotunneling of quasiparticles
through the box{]}. Such major facets are hard (if not impossible)
to access otherwise, e.g., using multiple Josephson periodicities
\cite{Lindner2012,Cheng2012,Clarke2013,Vaezi2013,Zhang2014,Cheng2015,foot1},
zero-bias anomalies \cite{Clarke2014}, split conductance peaks due
to finite-size effects \cite{Barkeshli2014a}, or quantized conductance
measurements \cite{Kim2017,foot1}. Apart from being interesting in
its own right, e.g., in the context of topological Kondo effects \cite{Beri2012},
it stands to reason that the experimental implementation of the PF
box would pave the way for realizing PF-based quantum information
devices \cite{foot_control-by-meas}.

\emph{PF box model.} For concreteness, we study an array of PF zero
modes implemented via two $\nu=1/(2k+1)$ FQH puddles of opposite
spin \cite{Lindner2012,Clarke2013,foot_Heiblum}, cf.~Fig.~\ref{fig1}.
Our proposal is adaptable to other platforms (see the Supplemental
Matterial \cite{SupplMat}); for $\nu=1$ it is reduced to a Majorana
setup \cite{Plugge2017}. The puddle edges are described by bosonic
fields $\hat{\phi}_{s=\uparrow/\downarrow=\pm1}(x)$ with commutator
\cite{Clarke2013}
\begin{equation}
[\hat{\phi}_{s}(x),\hat{\phi}_{s}(x')]=is\pi\sgn(x-x'),\quad[\hat{\phi}_{\uparrow}(x),\hat{\phi}_{\downarrow}(x')]=i\pi.\label{eq:phi_commutation}
\end{equation}
The resulting fractional helical edge state can be gapped by proximity
coupling to SC or FM segments, see Fig.~\ref{fig1}(a), with the
Hamiltonian $H_{0}=H_{\mathrm{edge}}+H_{\mathrm{SC}}+H_{\mathrm{FM}}+H_{\mathrm{C}}$,
where $H_{\mathrm{edge}}=(v/4\pi)\sum_{s}\int_{-\infty}^{\infty}dx(\partial_{x}\hat{\phi}_{s})^{2}$
with edge velocity $v$. Furthermore,
\begin{eqnarray}
H_{\mathrm{SC}} & = & -\Delta\sum_{j=1}^{N}\int_{\mathrm{SC}_{j}}dx\cos\left(\frac{\hat{\phi}_{\uparrow}(x)+\hat{\phi}_{\downarrow}(x)}{\sqrt{\nu}}+\hat{\varphi}\right),\label{eq:H_SC}\\
H_{\mathrm{FM}} & = & -t\sum_{j=1}^{N+1}\int_{\mathrm{FM}_{j}}dx\cos\left(\frac{\hat{\phi}_{\uparrow}(x)-\hat{\phi}_{\downarrow}(x)}{\sqrt{\nu}}\right).\label{eq:H_FM}
\end{eqnarray}
$H_{\mathrm{SC}}$ describes the SC pairing induced in the edges by
the proximitizing SCs, where $\hat{\varphi}$ is the SC phase operator
and $\Delta$ is the absolute value of the induced pairing amplitude.
$H_{\mathrm{FM}}$ describes electron hopping between the edges accompanied
by a spin flip, which is enabled by the presence of the FM. The hopping
amplitude $t$ is proportional to the FM in-plane magnetization caused
by, e.g., geometrical effects. All the proximitizing SCs (FMs) are
implied to be parts of one common SC (FM), see Fig.~\ref{fig1}(b).
For a floating (not grounded) SC, the charging term is $H_{\mathrm{C}}=(\hat{Q}_{0}-q_{\mathrm{g}})^{2}/(2C_{\mathrm{SC}})$,
with the charge $\hat{Q}_{0}$ satisfying $[\hat{\varphi},\hat{Q}_{0}]=2i$.
The offset charge $q_{\mathrm{g}}$ is controlled by a gate {[}Fig.~\ref{fig1}(b){]}.
Finally, the charge and spin of an edge segment are given by
\begin{equation}
\left.\begin{array}{c}
\hat{Q}_{AB}\\
\hat{S}_{AB}
\end{array}\right\} =\int_{A}^{B}dx(\hat{\rho}_{\uparrow}\pm\hat{\rho}_{\downarrow}),\quad\hat{\rho}_{s}(x)=\frac{s\sqrt{\nu}}{2\pi}\partial_{x}\hat{\phi}_{s}.\label{csseg}
\end{equation}
A parafermion $N$-box is then defined by $N+1$ FM and $N$ SC domains
excluding the outer edges. For instance, Fig.~\ref{fig1} shows a
two-box.

\emph{Low-energy theory.} The quasiparticle excitations in the SC
and FM domains have gaps $(4\pi v\Delta/\nu)^{1/2}$ and $(4\pi vt/\nu)^{1/2}$,
respectively \cite{Lindner2012}. At energies below these scales \cite{foot:qp_poisoning},
the problem can be simplified using the method of Ref.~\cite{Ganeshan2016}
since the large cosines in Eqs.~\eqref{eq:H_SC} and \eqref{eq:H_FM}
imply that each FM (SC) domain is then effectively described by an
integer-valued operator $\hat{m}_{j}$ ($\hat{n}_{j}$), see Refs.~\cite{Lindner2012,Clarke2013}
and Fig.~\ref{fig1}(a),
\begin{equation}
\left.\frac{\hat{\phi}_{\uparrow}(x)\mp\hat{\phi}_{\downarrow}(x)}{2\pi\sqrt{\nu}}\right|_{x\in\mathrm{FM}_{j}/\mathrm{SC}_{j}}=\left\{ \begin{array}{c}
\hat{m}_{j},\\
\hat{n}_{j}-\hat{\varphi}/(2\pi).
\end{array}\right.
\end{equation}
The only nontrivial commutation relation is $[\hat{m}_{j},\hat{n}_{l}]=i/(\pi\nu)$
for $j>l$, whereas $[\hat{m}_{j},\hat{n}_{l}]=0$ for $j\le l$.
Using Eq.~\eqref{csseg}, the charge $\hat{Q}_{j}$ (spin $\hat{S}_{j}$)
of the edge segment corresponding to $\mathrm{SC}_{j}$ ($\mathrm{FM}_{j}$
except for the first and the last FM) is
\begin{equation}
\hat{Q}_{j}=\nu(\hat{m}_{j+1}-\hat{m}_{j}),\quad\hat{S}_{j}=\nu(\hat{n}_{j}-\hat{n}_{j-1}).\label{eq:FM_spin}
\end{equation}
Note that FM (SC) domains cannot host charge (spin) at low energies.
The semi-infinite outer edges are merged with each other and decouple
from the PF box. Below, we probe the system by fractional quasiparticle
tunneling. At low energies, this can only happen at interfaces between
different domains. The projected low-energy quasiparticle operators
are (cf.~Refs.~\cite{Lindner2012,Clarke2013})
\begin{equation}
\hat{\alpha}_{j,s}=\left\{ \begin{array}{cc}
e^{i\pi\nu(\hat{n}_{l}+s\hat{m}_{l}-\hat{\varphi}/2\pi)}, & j=2l-1,\\[0.1cm]
e^{i\pi\nu(\hat{n}_{l}+s\hat{m}_{l+1}-\hat{\varphi}/2\pi)}, & j=2l,
\end{array}\right.\label{pfopdef}
\end{equation}
where $j$ is the domain-wall number and $s$ is the spin of the edge,
see Fig.~\ref{fig1}(a). The PF operators in Eq.~\eqref{pfopdef}
satisfy a $\mathbb{Z}_{n}$ parafermion algebra with index $n=2/\nu$
\cite{Alicea2016},
\begin{equation}
\hat{\alpha}_{j,s}\hat{\alpha}_{l,s}=\omega_{s}^{{\rm sgn}(l-j)}\hat{\alpha}_{l,s}\hat{\alpha}_{j,s},\quad\omega_{s}=e^{2\pi is/n}=e^{i\pi\nu s}.\label{parafermionalgebra}
\end{equation}

The low-energy Hilbert space of the box is now spanned by $|Q_{\mathrm{tot}},Q_{j=1,...,N-1}({\rm mod}~2)\rangle$,
where $\hat{Q}_{\mathrm{tot}}=\sum_{j=0}^{N}\hat{Q}_{j}=\hat{Q}_{0}+\nu(\hat{m}_{N+1}-\hat{m}_{1})$
is the total charge of the proximitizing SC and the FQH edges within
the PF box. Note that $Q_{\mathrm{tot}}$ has fractional values differing
by multiples of $\nu$. Since the SC can absorb electron pairs, the
remaining quantum numbers describe the distribution of fractional
quasiparticles over the SC domains of the PF box. The box Hamiltonian
is then given by
\begin{equation}
H_{\mathrm{box}}=\frac{1}{2C_{\mathrm{box}}}\left(\hat{Q}_{\mathrm{tot}}-q_{\mathrm{g}}\right)^{2}=E_{\mathrm{C}}\left(\frac{\hat{Q}_{\mathrm{tot}}-q_{\mathrm{g}}}{\nu}\right)^{2},\label{eq:H_box}
\end{equation}
where $C_{\mathrm{box}}$ is the effective box capacitance and all
states with the same $Q_{\mathrm{tot}}$ are degenerate up to exponentially
small splittings (see the Supplemental Material \cite{SupplMat}),
neglected here. Below we consider the simplest cases: one-box and
two-box. The Hilbert space of the one-box is spanned by $|Q_{\mathrm{tot}}\rangle$
and does not allow for a degenerate subspace. In contrast, for every
value of $Q_{\mathrm{tot}}$, the two-box has topological degeneracy
$n=2/\nu$ due to the different ways to distribute charge between
$\mathrm{SC}_{1}$ and $\mathrm{SC}_{2}$. A simple estimate puts
$E_{\mathrm{C}}$ in the range of 0.1–1\,K.

\begin{figure}
\centering \includegraphics[width=1\columnwidth]{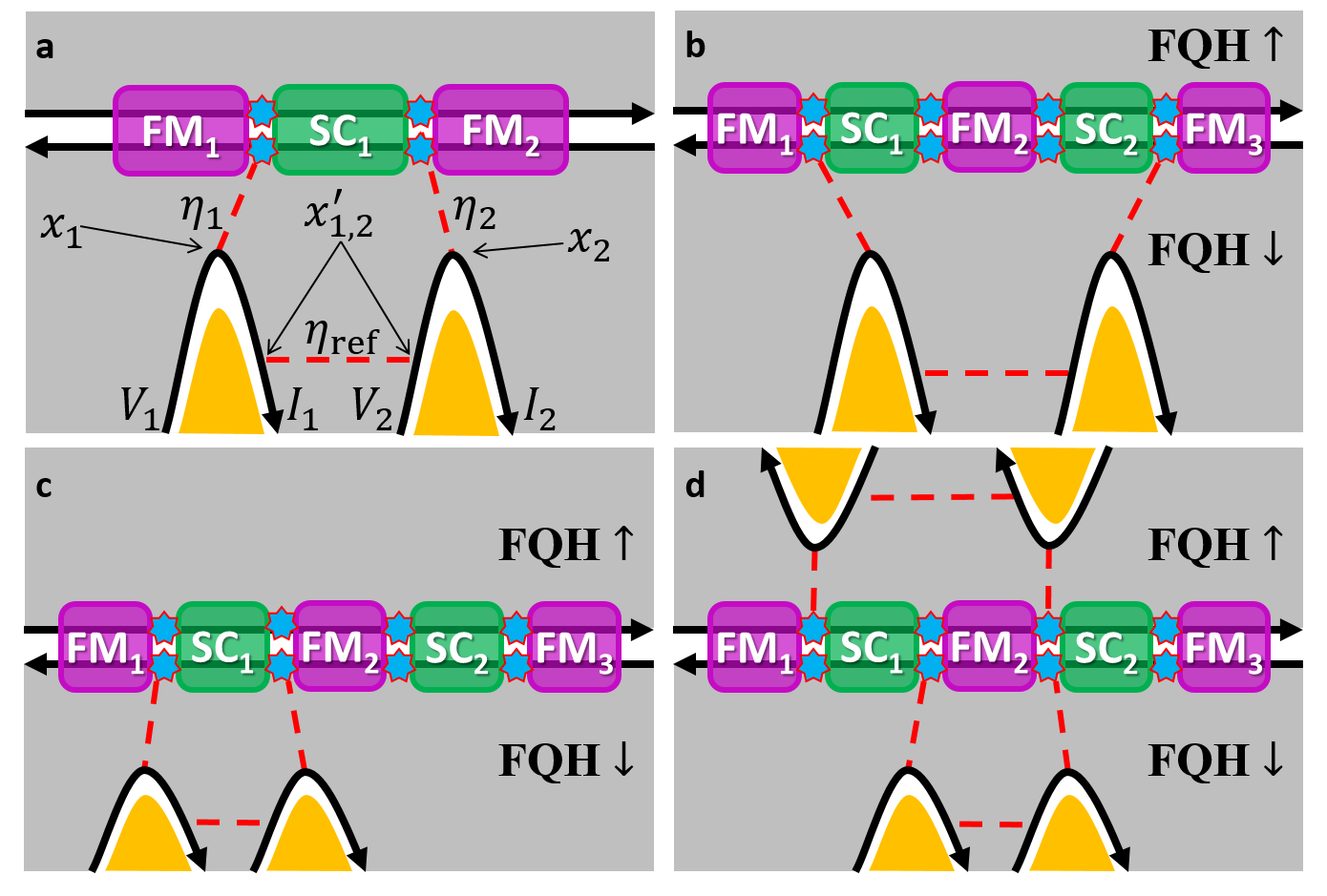} \caption{\label{fig2} Setups for measuring the PF box state employing additional
leads (FQH edges; the curved black arrows), tunnel couplings (the
dashed red lines), and gates (golden) controlling the distance between
the leads and the parafermions (and thus the strength of the tunnel
couplings). (a) Measuring $\hat{Q}_{\mathrm{tot}}~\mathrm{mod}~2$
in a one-box. (b,c,d) Setups for measuring various observables in
a two-box: (b) $(\hat{Q}_{\mathrm{tot}}-\hat{S}_{2})~\mathrm{mod}~2$;
(c) $\hat{Q}_{1}~\mathrm{mod}~2$; (d) $(\hat{S}_{2}-\hat{Q}_{1})~\mathrm{mod}~2$
and $\hat{S}_{2}~\mathrm{mod}~2$ for the upper and lower pairs of
leads, respectively.}
\end{figure}

\emph{Cotunneling Hamiltonian with FQH edges.} We next consider two
additional edge segments ($\gamma=1,2$), each approaching (near $x=x_{\gamma}$)
a PF zero mode on the box. Such edges serve as leads and correspond
to fields $\hat{\phi}_{\gamma}(x)$ with $[\hat{\phi}_{\gamma}(x),\hat{\phi}_{\gamma}(x')]=i\chi\pi\sgn(x-x')$,
where $\chi=\pm1$ is the edge chirality. With applied voltage $V_{\gamma}$,
the edge Hamiltonian is given by $H_{\gamma}=H_{{\rm edge}}[\hat{\phi}_{\gamma}]-\frac{\chi\sqrt{\nu}}{2\pi}V_{\gamma}\int dx\partial_{x}\hat{\phi}_{\gamma}$.
Quasiparticles can then tunnel with amplitude $\eta_{\gamma}$ between
the edge and the PF $\hat{\alpha}_{j_{\gamma},s}$, which is modeled
by the tunneling Hamiltonian $H_{\mathrm{T}}=\sum_{\gamma=1,2}\eta_{\gamma}e^{i\sqrt{\nu}\hat{\phi}_{\gamma}(x_{\gamma})}\hat{\alpha}_{j_{\gamma},s}^{\dagger}+\mathrm{H.c.},$
see Fig.~\ref{fig2}. On top of that, we allow for direct quasiparticle
tunneling between the two edges described by $H_{\mathrm{T}}^{\mathrm{ref}}=\eta_{\mathrm{ref}}e^{-i\sqrt{\nu}\hat{\phi}_{2}(x_{2}')}e^{i\sqrt{\nu}\hat{\phi}_{1}(x_{1}')}+\mathrm{H.c.}$
Although due to the large length of SC/FM domains, the points of tunneling
to PFs $x_{1,2}$ must differ from $x_{1,2}'$, in what follows we
put $x_{\gamma}'=x_{\gamma}$ for simplicity, cf. after Eq.\,(\ref{eq:measurement_current}).
Note that in all the above processes quasiparticles tunnel through
the FQH bulk.

We now assume that the box charging energy in Eq.~(\ref{eq:H_box})
is the largest energy scale. Away from Coulomb resonances $q_{\mathrm{g}}\ne\nu(\mathbb{Z}+1/2)$,
transport between two leads is then dominated by cotunneling. Performing
a Schrieffer–Wolff transformation and projecting to the state of $Q_{\mathrm{tot}}$
that minimizes the charging energy, $H_{\mathrm{box}}+H_{\mathrm{T}}\rightarrow H_{\mathrm{cot}}$,
we obtain to the leading order,
\begin{equation}
H_{\mathrm{cot}}=\eta_{\mathrm{cot}}\hat{\alpha}_{j_{2},s}^{\thinspace}\hat{\alpha}_{j_{1},s}^{\dagger}e^{-i\sqrt{\nu}\hat{\phi}_{2}(x_{2})}e^{i\sqrt{\nu}\hat{\phi}_{1}(x_{1})}+\mathrm{H.c.},\label{eq:H_cotunneling}
\end{equation}
where $\eta_{\mathrm{cot}}=-\eta_{1}\eta_{2}^{*}\left(U_{+}^{-1}+U_{-}^{-1}\right)$
with $U_{\pm}\sim E_{C}$ being the charging energy penalty for adding/removing
one fractional quasiparticle to/from the box. The total Hamiltonian
for transfer of quasiparticles between the leads is then given by

\begin{gather}
H_{\mathrm{T12}}=H_{\mathrm{T}}^{\mathrm{ref}}+H_{\mathrm{cot}}=\hat{\eta}_{\mathrm{T12}}e^{-i\sqrt{\nu}\hat{\phi}_{2}(x_{2})}e^{i\sqrt{\nu}\hat{\phi}_{1}(x_{1})}+\mathrm{H.c.},\nonumber \\
\hat{\eta}_{\mathrm{T12}}=\eta_{\mathrm{ref}}+\eta_{\mathrm{cot}}\hat{\alpha}_{j_{2},s}^{\thinspace}\hat{\alpha}_{j_{1},s}^{\dagger}.\label{eq:H_cotunneling_total}
\end{gather}
Since the operator $\hat{\alpha}_{j_{2},s}^{\thinspace}\hat{\alpha}_{j_{1},s}^{\dagger}$
in Eq.~\eqref{eq:H_cotunneling_total} acts only in the discrete
box subspace, the effective cotunneling Hamiltonian $H_{\mathrm{T}12}$
corresponds to quasiparticle tunneling between the leads with effective
amplitude $\hat{\eta}_{\mathrm{T12}}$. Noting that $(\hat{\alpha}_{j_{2},s}^{\thinspace}\hat{\alpha}_{j_{1},s}^{\dagger})^{n}=-1$
with $n=2/\nu$, one sees that the properties specific to PFs are
encoded by the cotunneling phase. Indeed, the eigenvalues of $\hat{\alpha}{}_{j_{2},s}^{\thinspace}\hat{\alpha}_{j_{1},s}^{\dagger}$
follow as $e^{i\pi\nu(r+1/2)}$ (integer $r$), and the cotunneling
phase therefore depends on the PF box state with possible phase values
differing by multiples of $\pi\nu$.

\emph{Cotunneling current.} Next we observe that $H_{\mathrm{T}12}$
is relevant under the renormalization group (RG) with scaling dimension
equal to $\nu<1$ \cite{foot:scaling_dimension}. The RG flow towards
the strong quasiparticle tunneling regime eventually implies a two-terminal
conductance $\nu e^{2}/h$ \cite{Kim2017}. However, for a finite
voltage $V_{12}=V_{1}-V_{2}$ with $|V_{12}|\gg V_{B}\propto|\eta_{\mathrm{T12}}|^{1/(1-\nu)}$,
the RG flow is effectively cut off. For a given $V_{12}$, this inequality
is always realized for sufficiently small tunnel couplings. The tunneling
current $I_{\mathrm{T}\mathrm{12}}=I_{2}-\nu e^{2}V_{2}/h$ between
the two leads then follows from perturbation theory in $H_{\mathrm{T12}}$
\cite{Wen1991},
\begin{equation}
\hat{I}_{\mathrm{T}\mathrm{12}}=\frac{|\hat{\eta}_{\mathrm{T12}}|^{2}}{v^{2\nu}}\frac{2\pi\nu}{\Gamma(2\nu)}(\nu|V_{12}|)^{2\nu-1}\sgn(V_{12}),\label{eq:measurement_current}
\end{equation}
$\Gamma(x)$ being the Euler gamma function. For a box initially in
a superposition of different $\hat{\alpha}_{j_{2},s}^{\thinspace}\hat{\alpha}_{j_{1},s}^{\dagger}$
eigenstates, such a current measurement implies a projection to the
observed eigenstate, cf.~Refs.~\cite{Landau2016,Plugge2017}. By
measuring $\hat{I}_{\mathrm{T}\mathrm{12}}$ and hence $|\hat{\eta}_{\mathrm{T12}}|$,
one can therefore characterize the quantum state of a PF box. In this
calculation we assumed low temperatures ($T\ll|V_{12}|$) and long
edges ($v/L\ll|V_{12}|$). Furthermore, we have put the points of
direct tunneling between the two edges ($x_{1,2}'$) to coincide with
the points for tunneling to parafermions ($x_{1,2}$), cf. Fig.~\ref{fig2}(a).
Realistically, they will not coincide, resulting in a suppression
of the interference term $\sim\eta_{\mathrm{cot}}\eta_{\mathrm{ref}}^{*}\hat{\alpha}_{j_{2},s}^{\thinspace}\hat{\alpha}_{j_{1},s}^{\dagger}+\mathrm{H.c.}$
which enables one to measure the eigenvalue of $\hat{\alpha}_{j_{2},s}^{\thinspace}\hat{\alpha}_{j_{1},s}^{\dagger}$.
This suppression reduces (but does not destroy) the interference visibility.

\begin{figure}[t]
\centering \includegraphics[width=1\columnwidth]{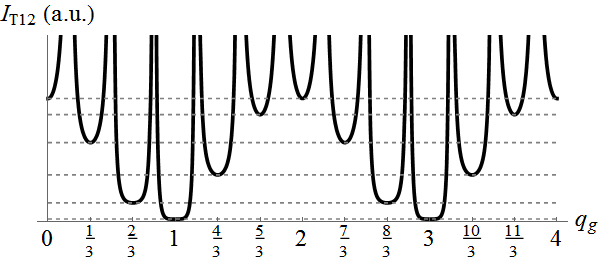} \caption{\label{fig3} Sketch of the tunneling current $I_{\mathrm{T}\mathrm{12}}$
vs gate parameter $q_{\mathrm{g}}$ for the one-box configuration
in Fig.~\ref{fig2}(a) with $\nu=1/3$. The current in a Coulomb
valley depends on $Q_{\mathrm{tot}}~\mathrm{mod}~2$ and thus on $q_{\mathrm{g}}$.
When gradually increasing $q_{\mathrm{g}}\rightarrow q_{\mathrm{g}}+2$,
one passes through $n=2/\nu$ peaks with much stronger (yet finite)
current. At the end of this sequence, the same current $I_{\mathrm{T}\mathrm{12}}$
is repeated.}
\end{figure}

\emph{Number of eigenvalues.} Consider the configuration in Fig.~\ref{fig2}(a),
which gives access to the operator $\hat{\alpha}_{2,\downarrow}^{\thinspace}\hat{\alpha}_{1,\downarrow}^{\dagger}=e^{-i\pi(\hat{Q}_{\mathrm{tot}}+\nu/2)}$,
cf.~Eqs.~\eqref{pfopdef} and \eqref{eq:H_cotunneling_total}, via
two leads connected to a one-box. We now tune the box to a Coulomb-blockade
valley. By varying the gate voltage across a peak $q_{\mathrm{g}}\rightarrow q_{\mathrm{g}}+\nu$,
we also enforce $Q_{\mathrm{tot}}\rightarrow Q_{\mathrm{tot}}+\nu$
since the box will adhere to the ground state of $H_{\mathrm{box}}$.
As a consequence, we effectively obtain $\eta_{\mathrm{cot}}\rightarrow\eta_{\mathrm{cot}}e^{-i\pi\nu}$
in Eq.~\eqref{eq:H_cotunneling_total} and thus a different tunneling
current $I_{\mathrm{T}\mathrm{12}}$. After switching through $n=2/\nu$
subsequent Coulomb valleys, we have $q_{\mathrm{g}}\rightarrow q_{\mathrm{g}}+2$
and return to the original value of $I_{\mathrm{T}\mathrm{12}}$.
This characteristic dependence of $I_{\mathrm{T}\mathrm{12}}$ on
$q_{\mathrm{g}}$ is sketched in Fig.~\ref{fig3}. Such an experiment
can already determine the number $n$ of possible eigenvalues of $\hat{\alpha}_{2,\downarrow}^{\thinspace}\hat{\alpha}_{1,\downarrow}^{\dagger}$,
which is a nontrivial check of the PF operator properties.

\emph{Degenerate PF space.} Next we show how to engineer an $n$-fold
degenerate PF space. To accomplish this task, we consider a two-box
as shown in Figs.~\ref{fig2}(b)–\ref{fig2}(d). The two-box has
two degrees of freedom, namely, $Q_{\mathrm{tot}}$, which behaves
the same way as for the one-box, and $Q_{1}~(\mathrm{mod}~2)$, which
labels the $n$-fold degenerate subspace for any given $Q_{\mathrm{tot}}$.
The two-box contains four domain walls, allowing for various options
to access PF operator combinations. For instance, by connecting leads
to the first and the last domain walls, cf.~Fig.~\ref{fig2}(b),
a current measurement determines the phase of $\hat{\alpha}_{4,s}^{\thinspace}\hat{\alpha}_{1,s}^{\dagger}=e^{i\pi(s\hat{Q}_{\mathrm{tot}}+s\nu/2+\hat{S}_{2})}$.
Repeating the above one-box protocol then implies a $q_{\mathrm{g}}$-dependent
behavior as in Fig.~\ref{fig3}. Next, we note that the subspace
of fixed $Q_{\mathrm{tot}}$ is spanned by the $|Q_{1}~\mathrm{mod}~2\rangle$
states. These states correspond to the eigenvalues $e^{is\pi(Q_{1}+\nu/2)}$
of $\hat{\alpha}_{2,s}^{\thinspace}\hat{\alpha}_{1,s}^{\dagger}$,
which in turn follow from current measurements as shown in Fig.~\ref{fig2}(c);
$\hat{\alpha}_{3,s}^{\thinspace}\hat{\alpha}_{2,s}^{\dagger}=e^{i\pi(\hat{S}_{2}+s\nu/2)}$
and $\hat{\alpha}_{3,s}^{\thinspace}\hat{\alpha}_{1,s}^{\dagger}=e^{i\pi(s\hat{Q}_{1}+\hat{S}_{2}+s\nu/2)}$
implying that this degenerate subspace is also spanned by the eigenbasis
of $\hat{S}_{2}$ or of $\hat{S}_{2}\pm\hat{Q}_{1}$. Both operators
can be accessed as illustrated in Fig.~\ref{fig2}(d). Now let us
take any eigenstate $|Q_{1}~\mathrm{mod}~2\rangle$ of $\hat{\alpha}_{2,s}^{\thinspace}\hat{\alpha}_{1,s}^{\dagger}$.
Decomposing it into the eigenstates of $\hat{\alpha}_{3,s}^{\thinspace}\hat{\alpha}_{2,s}^{\dagger}$,
we obtain
\begin{equation}
|Q_{1}~\mathrm{mod}~2\rangle=\frac{1}{\sqrt{n}}\sum_{r=0}^{n-1}e^{i\pi\lambda_{r}}|(S_{2}~\mathrm{mod}~2)=\nu r\rangle,\label{eq:Q1_S2_decomposition}
\end{equation}
where $\lambda_{r}=r(Q_{1}~\mathrm{mod}~2)$. Similar statements hold
for any pair of the above observables. This structure allows one to
confirm the degeneracy of the subspace.

Indeed, let us select an arbitrary pair of noncommuting bilinear PF
operators, e.g., $\hat{O}_{1}=\hat{\alpha}_{2,\uparrow}^{\thinspace}\hat{\alpha}_{1,\uparrow}^{\dagger}$
and $\hat{O}_{2}=\hat{\alpha}_{3,\downarrow}^{\thinspace}\hat{\alpha}_{2,\downarrow}^{\dagger}$.
A measurement of $\hat{O}_{1}$ now projects the box state onto one
of its $n$ eigenstates. Then one measures $\hat{O}_{2}$, which should
project the system with equal probabilities, see Eq.~\eqref{eq:Q1_S2_decomposition},
onto any eigenstate of $\hat{O}_{2}$. Similarly, a subsequent measurement
of $\hat{O}_{1}$ projects this state with equal probabilities onto
one of the eigenstates of $\hat{O}_{1}$. Repeating this procedure
many times, one can verify that $\hat{O}_{1}$ (or $\hat{O}_{2}$)
has precisely $n$ possible eigenvalues. To explicitly check the degeneracy
one may now proceed as follows. One first performs repetitive measurements
of $\hat{O}_{1}$ with arbitrary time intervals between consecutive
current measurements. If we always find the same eigenvalue, we know
that $\left[\hat{O}_{1},H_{\mathrm{box}}\right]=0$. Now repeat this
procedure for the operator $\hat{O}_{2}$, which does not commute
with $\hat{O}_{1}$. If we also find $\left[\hat{O}_{2},H_{\mathrm{box}}\right]=0$,
all eigenstates of both $\hat{O}_{1}$ and $\hat{O}_{2}$ have trivial
time evolution, which proves the degeneracy of the PF space. We emphasize
that checking the degeneracy and the size of the PF subspace is an
important and nontrivial validation of the system properties.

\begin{figure}[t]
\centering \includegraphics[width=1\columnwidth]{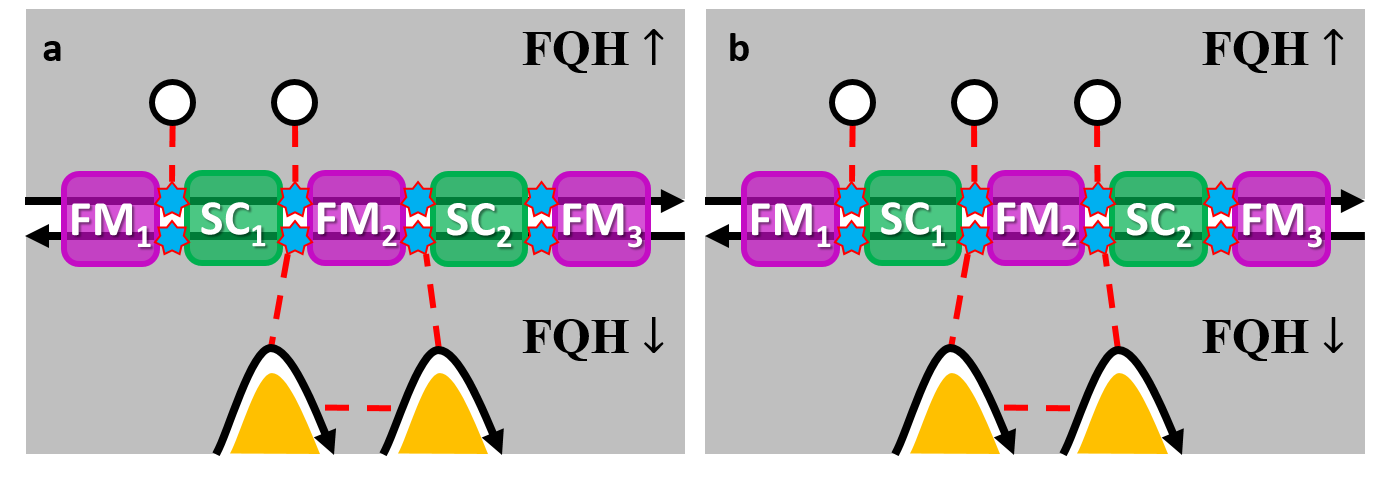} \caption{\label{fig4} Setups allowing for two-box manipulation with QADs (the
white circles). (a) The operator $\hat{\alpha}_{2,\uparrow}^{\thinspace}\hat{\alpha}_{1,\uparrow}^{\dagger}$
can be applied to an arbitrary PF box state by pumping a quasiparticle
between the QAD pair. In the same setup, $\hat{S}_{2}~\mathrm{mod}~2$
can be measured from the tunneling current between the shown leads.
(b) A setup allowing for application of operations $\hat{\alpha}_{2,\uparrow}^{\thinspace}\hat{\alpha}_{1,\uparrow}^{\dagger}$
and $\hat{\alpha}_{3,\uparrow}^{\thinspace}\hat{\alpha}_{2,\uparrow}^{\dagger}$,
which is sufficient to generate any digital operation that can be
applied through QADs, and measuring $\hat{S}_{2}~\mathrm{mod}~2$.
Gates used to create QADs are kept implicit.}
\end{figure}

\emph{PF state manipulation. }Finally we discuss how to perform on-demand
transitions in the $n$-fold degenerate subspace of a two-box with
QADs surrounded by the FQH liquid \cite{Kivelson1989,Kivelson1990,Goldman1996,Goldman1997,Maasilta1997}
as access units, see Fig.~\ref{fig4}. At low-energy scales, a QAD
in the Coulomb-blockade regime is equivalent to a two-level system,
\begin{equation}
H_{\mathrm{QAD}}=\nu V_{\mathrm{QAD}}\left(\hat{\psi}_{\mathrm{QAD}}^{\dagger}\hat{\psi}_{\mathrm{QAD}}^{\thinspace}-\frac{1}{2}\right)=\frac{V_{\mathrm{QAD}}}{n}\begin{pmatrix}1 & 0\\
0 & -1
\end{pmatrix},\label{eq:H_QAD}
\end{equation}
where $V_{\mathrm{QAD}}$ is an electrostatic gate potential and $\hat{\psi}_{\mathrm{QAD}}$
is the quasiparticle annihilation operator on the QAD. Consider now
two QADs coupled to the two-box as in Fig.~\ref{fig4}(a). Similar
to Eq.~\eqref{eq:H_cotunneling_total}, elastic cotunneling between
two QADs via the PF box is described by $H_{\mathrm{cot,QAD}}=\eta_{\mathrm{cot}}\hat{\psi}_{\mathrm{QAD2}}^{\dagger}\hat{\psi}_{\mathrm{QAD1}}^{\thinspace}\hat{\alpha}_{j_{2},s}^{\thinspace}\hat{\alpha}_{j_{1},s}^{\dagger}+\mathrm{H.c.}$,
where the amplitude $\eta_{\mathrm{cot}}$ now does not renormalize
anymore. By adiabatically pumping a quasiparticle from QAD1 to QAD2
through suitable changes in the gate voltages $V_{\mathrm{QAD1/2}}$,
an arbitrary PF box state $|\Phi\rangle$ must then transform according
to $|\Phi\rangle\rightarrow\hat{\alpha}_{j_{2},s}^{\thinspace}\hat{\alpha}_{j_{1},s}^{\dagger}|\Phi\rangle$.
QADs thus facilitate digital operations $\hat{\alpha}_{j,s}^{\thinspace}\hat{\alpha}_{l,s}^{\dagger}$
within the degenerate PF subspace. Furthermore, employing protocols
for both measurement with leads and manipulation with QADs, one can
provide direct manifestations of the PF algebra \eqref{parafermionalgebra};
e.g., by measuring $\hat{\alpha}_{3,\downarrow}^{\thinspace}\hat{\alpha}_{2,\downarrow}^{\dagger}=e^{i\pi(\hat{S}_{2}-\nu/2)}$
and applying $\hat{\alpha}_{2,\uparrow}^{\thinspace}\hat{\alpha}_{1,\uparrow}^{\dagger}=e^{i\pi(\hat{Q}_{1}+\nu/2)}$
which shifts $\hat{S}_{2}\rightarrow\hat{S}_{2}-\nu~(\mathrm{mod}~2)$,
cf. Fig.~\ref{fig4}(a). We emphasize that all nontrivial digital
operations are generated by two operators, e.g., $\hat{\alpha}_{2,\uparrow}^{\thinspace}\hat{\alpha}_{1,\uparrow}^{\dagger}$
and $\hat{\alpha}_{3,\uparrow}^{\thinspace}\hat{\alpha}_{2,\uparrow}^{\dagger}$.
These can be implemented with three QADs {[}Fig.~\ref{fig4}(b){]}.

\emph{Conclusions.} The parafermion box introduced in this Rapid Communication
can simplify and facilitate experimental studies of PF-based quantum
states. Our proposed measurement protocols, which employ fractional
edge states as leads and/or quantum antidots for state manipulation,
crucially rely on the unique and intrinsically nonlocal ways to access
the box in the Coulomb-blockade regime. One can thereby largely avoid
several difficulties that may affect earlier proposals for observing
PF physics. In particular, we have shown how to observe the dimension
of the zero-mode space, how to realize and demonstrate the existence
of a degenerate space, and how to perform digital operations in this
degenerate state manifold. The results of our protocols are distinctly
different from Coulomb-blockade signatures of anyonic tunneling and
should enable the experimental confirmation of the parafermion algebra
in Eq.~\eqref{parafermionalgebra}.

\begin{acknowledgments}

\emph{Acknowledgments}. We thank A. Altland, K. Flensberg, L. Glazman,
and S. Plugge for discussions. We acknowledge funding by the Deutsche
Forschungsgemeinschaft (Bonn) within the network CRC TR 183 (Project
No. C01) and Grant No.~RO 2247/8-1, by the IMOS Israel-Russia program,
by the ISF, and the Italia-Israel Project QUANTRA. This Rapid Communication
was prepared with the help of the LyX software \cite{LyX}.

\end{acknowledgments}

\clearpage


\section*{Supplementary material}

# Supplemental Material to "Measurement and Control of Coulomb-Blockaded Parafermion Box"

Kyrylo Snizhko,[1] Reinhold Egger,[2] and Yuval Gefen[1]

[1]*Department of Condensed Matter Physics, Weizmann Institute of Science, Rehovot, 76100 Israel*
[2]*Institut für Theoretische Physik, Heinrich-Heine-Universität, D-40225 Düsseldorf, Germany*

(Dated: February 17, 2018)

In this Supplemental Material we discuss some technical aspects of our model and the applicability of our results to the $\nu = 2/3$ platform for parafermions. We discuss the issue of whether tuning momenta is necessary to open a gap in both FM and SC domains in Sec. I. Sec. II discusses how our model for the parafermion box changes when there are order parameter phase differences between the SC domains, and how these phase differences can be usefully utilized. Sec. III provides technical details regarding the derivation of the box charging energy. In Sec. IV, we elaborate on how the two-lead measurement scheme implements a projective measurement. In Sec. V, the relation between the eigenbases of different observables acting in the 2-box degenerate subspace is derived. Sec. VI elaborates on the protocol for manipulating an $N$-box degenerate subspace via QADs. Finally, Sec. VII discusses the applicability of our results to parafermions generated on the boundaries of $\nu = 2/3$ puddles.

## I. WHY NO MOMENTA TUNING IS NECESSARY IN FM/SC DOMAINS

The issue of whether a gap can be opened simultaneously at FM and SC domains arises due to the following naive argument. The tunneling, which is uniform along the edge, conserves momentum and therefore opening a gap in the FM domain requires the two edges to have the same Fermi momenta. Similarly, a conventional SC proximitizing the system induces pairing at opposite momenta, and opening a gap in the SC domains requires the Fermi momenta of the edges to be opposite. Therefore, according to the argument, the gap cannot be opened simultaneously in FM and SC domains unless the Fermi momenta of both edges are at zero. The resolution comes from the fact that the system is subject to a magnetic field. For a particle in magnetic field there is a difference between the canonical momentum $\hat{\mathbf{p}}$ and the physical (kinetic) momentum $\hat{\pi} = \hat{\mathbf{p}} + \mathbf{A}$ [1]. Due to translation invariance, the tunneling conserves the canonical momentum, while the SC pairing happens at opposite physical momenta (since the energies of the paired electrons $E = \hat{\pi}^2/2m$ in the proximitizing SC should be the same).

Qualitatively, one can argue that both conditions can be satisfied simultaneously by the following consideration. Consider a free electron in a plane subjected to a perpendicular magnetic field, using the Landau gauge $\mathbf{A} = (By, 0)$. In an eigenstate of the Hamiltonian, the particle's canonical momentum is related to its average position as $p_x = -By = -A_x(y)$. Therefore, two close-by FQH edges have the same canonical Fermi momentum. The physical momentum can be defined as $\pi_x = m\partial E/\partial p_x = \pm mv$, where we used the energy at the edge $E = \pm v(p_x - p_F)$, the sign $\pm$ denotes opposite propagation directions of the two edges, and $v$ is the edge velocity. Therefore the condition for opposite physical momenta at the edges is satisfied automatically. Below we provide a more technical analysis of the issue which highlights a few subtleties.

For simplicity, we concentrate on the case of $\nu = 1$ and discuss the issue in the fermionic language. This has the advantage that one can describe the two integer quantum Hall (IQH) edges by solving the problem of a particle in the magnetic field with a potential barrier which separates the plane along the line $y = y_0$. Consider a narrow high barrier that separates the two IQH puddles. Then electrons in each edge are described by the field operator

$$\hat{\psi}_s(x) = \int_{-\Lambda}^{\Lambda} \frac{dk}{\sqrt{2\pi}} \hat{a}_{ks} e^{ikx}, \tag{S1}$$

$$\left\{\hat{a}_{ks}, \hat{a}_{k's'}^\dagger\right\} = \delta_{kk'}\delta_{ss'}, \tag{S2}$$

where $s = \uparrow/\downarrow = \pm 1$ and $\Lambda$ is a cutoff. The edge Hamiltonian is given by

$$H_{\text{edge}} = \int_{-\Lambda}^{\Lambda} dk \sum_s \varepsilon_s(k) \hat{a}_{ks}^\dagger \hat{a}_{ks}, \tag{S3}$$

$$\varepsilon_s(k) = sv(k - k_s^F), \tag{S4}$$

where $k_s^F$ is the Fermi momentum. In order for this low-energy description to be valid and $k_s^F \in (-\Lambda, \Lambda)$, the barrier separating the two edges should be much thinner than the magnetic length $l_B = B^{-1/2}$.

Electron tunneling in a FM domain can be described by

$$H_{\text{FM}} = \tilde{t} \int_{-\infty}^{+\infty} dx \hat{\psi}_\uparrow^\dagger(x) \hat{\psi}_\downarrow(x) + \text{h.c.}$$
$$= \tilde{t} \int dk \hat{a}_{k\uparrow}^\dagger \hat{a}_{k\downarrow} + \text{h.c.} \tag{S5}$$

Diagonalising $H = H_{\text{edge}} + H_{\text{FM}}$ one finds the following

dispersion relation

$$\varepsilon_{\pm}(k) = v\frac{k_{\downarrow}^F - k_{\uparrow}^F}{2} \pm \sqrt{|\tilde{t}|^2 + v^2\left(k - \frac{k_{\downarrow}^F + k_{\uparrow}^F}{2}\right)^2}. \quad \text{(S6)}$$

Comparing this to the Fermi level (which is located at zero energy), one finds that the excitation gap is equal to $\min\left(|\tilde{t}| - \frac{v}{2}|k_{\downarrow}^F - k_{\uparrow}^F|, 0\right)$. As can be seen from Eq. (S4), tuning chemical potential is equivalent to shifting $k_{\uparrow}^F$ and $k_{\downarrow}^F$ in opposite directions. Therefore, by tuning the edges' chemical potential one can maximize the excitation gap at $|\tilde{t}|$.

Consideration of a SC domain is somewhat more involved. The pairing is described by

$$H_{\text{SC}} = \int_{-\infty}^{+\infty} dx\tilde{\Delta}(x)\hat{\psi}_{\uparrow}^{\dagger}(x)\hat{\psi}_{\downarrow}^{\dagger}(x) + \text{h.c.}, \quad \text{(S7)}$$

$$\tilde{\Delta}(x) = \left\langle \hat{\psi}_{\uparrow,\text{SC}}(x, y_{\uparrow})\hat{\psi}_{\downarrow,\text{SC}}(x, y_{\downarrow})\right\rangle, \quad \text{(S8)}$$

where $\hat{\psi}_{s,\text{SC}}$ are electron operators in the proximitizing SC and $y_s$ denote the location of the IQH edges and are related to $k_s^F$. In the absence of the magnetic field,

$$\tilde{\Delta}(x) = \int d^3k \left\langle \hat{a}_{\mathbf{k},\uparrow,\text{SC}}\hat{a}_{-\mathbf{k},\downarrow,\text{SC}}\right\rangle e^{2ik_y(y_{\uparrow} - y_{\downarrow})}$$
$$= \tilde{\Delta}_0 f(y_{\uparrow} - y_{\downarrow}) \approx \tilde{\Delta}_0, \quad \text{(S9)}$$

where $f(y)$ is of order 1 when $|y|$ is smaller that the SC coherence length $\xi$. If one puts the same SC into a location where the vector potential is uniform and has the value $\mathbf{A}_0 = (A_{0x}, 0)$, all the canonical momenta get shifted by $\mathbf{A}_0$ resulting in

$$\tilde{\Delta}(x) = \int d^3k \left\langle \hat{a}_{\mathbf{k}-\mathbf{A}_0,\uparrow,\text{SC}}\hat{a}_{-\mathbf{k}-\mathbf{A}_0,\downarrow,\text{SC}}\right\rangle$$
$$\times e^{2ik_y(y_{\uparrow} - y_{\downarrow}) - 2iA_{0x}x} = \tilde{\Delta}_0 e^{-2iA_{0x}x}. \quad \text{(S10)}$$

In the presence of the magnetic field, i.e., when the vector potential $\mathbf{A}$ varies in space, the behavior of the SC order parameter $\tilde{\Delta}(x)$ may be modified significantly due to the SC screening the magnetic field. Realistically, the proximitizing SC will be a type-2 SC penetrated by vortices. Away from vortices, one can approximately use the previous expression for $\tilde{\Delta}(x)$. Taking into account the phase acquired by the Cooper pair electrons while travelling from the proximitizing SC to the edges, we write

$$\tilde{\Delta}(x) = \tilde{\Delta}_0 e^{-i(A_x(y_{\uparrow}) + A_x(y_{\downarrow}))x}. \quad \text{(S11)}$$

Solving $H = H_{\text{edge}} + H_{\text{SC}}$ with this $\tilde{\Delta}(x)$ we find the excitation gap to be $\min\left(|\tilde{\Delta}_0| - \frac{v}{2}|k_{\downarrow}^F + k_{\uparrow}^F + A_x(y_{\uparrow}) + A_x(y_{\downarrow})|, 0\right)$. For a free particle $k_x + A_x(y) = 0$; however, the presence of the barrier separating the edges makes $k_s^F$ deviate from $-A_x(y_s)$. If the barrier is symmetric, one can show that $k_{\downarrow}^F + k_{\uparrow}^F + A_x(y_{\uparrow}) + A_x(y_{\downarrow}) = 0$. In the case when the barrier is asymmetric, one expects the deviation from zero to be a fraction of $Bl_B$. For $B = 10\,\text{T}$ and $v = 10^4\,\text{m/s}$ we estimate $vBl_B/2 \approx 3\,\text{K}$. Therefore, the induced SC amplitude $\tilde{\Delta}_0 \approx 5\,\text{K}$ should be sufficient to create an observable gap in SC domains even for an asymmetric barrier separating the two quantum Hall puddles.

Therefore, in order to induce gaps in both FM and SC domains, one needs the two quantum Hall edges to be located within less than a magnetic length from each other in $y$ direction, the barrier separating them has to be sufficiently symmetric, and the chemical potential of the edges has to be tuned into the FM domains' gap. However, no momentum tuning is necessary. We note that the issues of the separating barrier and/or vortices in the proximitizing SC influencing the SC domains warrant a separate study.

A referee of the manuscript brought to our attention recent results [2] showing that in a realistic system the induced SC pairing amplitude may depend significantly on the distance between the edges, which is not captured by the model above. This may be an important factor for experimental implementation of the system.

## II. NON-UNIFORM SC ORDER PARAMETER

An additional control handle can be brought to the PF box setup discussed in the main text through enabling the order parameter of different SC domains to have different phases. This can be done via applying a magnetic field in the $y$ direction (cf. Fig. S1). The SC proximitizing Hamiltonian then becomes

$$H_{\text{SC}} = -\Delta \sum_{j=1}^{N} \int_{\text{SC}_j} dx \cos\left(\frac{\hat{\phi}_{\uparrow}(x) + \hat{\phi}_{\downarrow}(x)}{\sqrt{\nu}} + \hat{\varphi}_j\right), \quad \text{(S12)}$$

where $\hat{\varphi}_j = \hat{\varphi} + \varphi_j$, $\hat{\varphi}$ is the bulk SC phase operator, and $\{\varphi_j\}$ are the phase differences between the domains and the bulk SC due to the fact that the system is subject to the magnetic field $B_y$. As a result of this modification, the expressions for the corresponding box operators become:

$$\left.\frac{\hat{\phi}_{\uparrow}(x) \mp \hat{\phi}_{\downarrow}(x)}{2\pi\sqrt{\nu}}\right|_{x \in \text{FM}_j/\text{SC}_j} = \begin{cases} \hat{m}_j, \\ \hat{n}_j - \hat{\varphi}_j/(2\pi). \end{cases}, \quad \text{(S13)}$$

$$\hat{Q}_j = \nu(\hat{m}_{j+1} - \hat{m}_j), \quad \text{(S14)}$$

$$\hat{S}_j = \nu(\hat{n}_j - \hat{n}_{j-1}) - \nu(\hat{\varphi}_j - \hat{\varphi}_{j-1})/(2\pi), \quad \text{(S15)}$$

$$\hat{\alpha}_{j,s} = \begin{cases} e^{i\pi\nu(\hat{n}_l + s\hat{m}_l - \hat{\varphi}_l/2\pi)}, & j = 2l - 1, \\ e^{i\pi\nu(\hat{n}_l + s\hat{m}_{l+1} - \hat{\varphi}_l/2\pi)}, & j = 2l. \end{cases} \quad \text{(S16)}$$

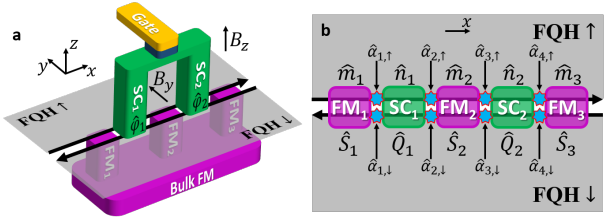

Figure S1. *The layout of a PF box with applied $B_y$ magnetic field*. a – The application of $B_y$ magnetic field creates phase difference between different SC domains. b – The physcial structure of the PF box remains the same as for $B_y = 0$. However, the expressions for operators $\hat{n}_j$, $\hat{S}_j$, and $\hat{\alpha}_{j,s}$ [cf. Eqs. (5-7)] are somewhat modified, see Eqs. (S13-S16).

See Eqs. (5-7) in the main text for corresponding expressions at $B_y = 0$.

Note that $(\hat{\alpha}_{j_2,s}\hat{\alpha}^\dagger_{j_1,s})^n = e^{i\delta\varphi}$ with $n = 2/\nu$ and $\delta\varphi = \pi + \varphi_{[(j_1+1)/2]} - \varphi_{[(j_2+1)/2]}$, where $[x]$ is the integer part of $x$. The eigenvalues of $\hat{\alpha}_{j_2,s}\hat{\alpha}^\dagger_{j_1,s}$ are then $e^{i\nu(\pi r + \delta\varphi/2)}$ (integer $r$). Therefore, the tunability of the phase differences $\varphi_j - \varphi_l$ is advantageous in the protocol for measuring the box operators proposed in the main text (cf. Fig. 2 and Eqs. (11-12)): whenever the two parafermionic operators $\hat{\alpha}_{j_1,s}$, $\hat{\alpha}_{j_2,s}$ belong to different SC domains, one may control the interference phase offset in $\hat{\eta}_{T12} = \eta_{\mathrm{ref}} + \eta_{\mathrm{cot}}\hat{\alpha}_{j_2,s}\hat{\alpha}^\dagger_{j_1,s}$ in order to produce optimal conditions for distinguishing different box states. Note that while a similar effect can be achieved by varying $B_z$ and thus adjusting directly the phase difference between $\eta_{\mathrm{ref}}$ and $\eta_{\mathrm{cot}}$ through Aharonov-Bohm effect, varying the perpendicular field would have the unwanted side effect of influencing the FQH liquids.

## III. THE CHARGING ENERGY OF THE PARAFERMIONIC BOXES

Here we provide technical details about the derivation of the low-energy theory of the PF box model, putting special emphasis on the role of the charging energy. In the limit $\Delta, t \to +\infty$ the fields tend to be pinned to the cosine minima, see Eqs. (2,3) of the main text. Expanding the cosines near their minima one can show for each domain that the finite momentum excitations in the domains are gapped (see, e.g., Appendix A1 of Ref. [3]). Therefore, at low energies the cosine arguments in Eqs. (2,3) of the main text are constant within the respective domains. In this approximation, imposing continuity of the fields $\hat{\phi}_s(x)$, one obtains for $\hat{\theta}_\pm(x) = \hat{\phi}_\uparrow(x) \pm \hat{\phi}_\downarrow(x)$

$$\hat{\theta}_+(x) = \begin{cases} 2\pi\sqrt{\nu}\hat{n}_j - \sqrt{\nu}\hat{\varphi} & , x \in \mathrm{SC}_j \\ \frac{2\pi(x - x_{2j-2})}{\sqrt{\nu}L_{\mathrm{FM}}}\hat{S}_j + \hat{\theta}_+(x^-_{2j-2}) & , x \in \mathrm{FM}_j, \end{cases} \quad \text{(S17)}$$

$$\hat{\theta}_-(x) = \begin{cases} 2\pi\sqrt{\nu}\hat{m}_j & , x \in \mathrm{FM}_j \\ \frac{2\pi(x - x_{2j-1})}{\sqrt{\nu}L_{\mathrm{SC}}}\hat{Q}_j + \hat{\theta}_-(x^-_{2j-1}) & , x \in \mathrm{SC}_j, \end{cases} \quad \text{(S18)}$$

where $L_{\mathrm{FM/SC}}$ is the length of the respective domain, $x_j$ is the coordinate of the $j$th domain wall ($x_0$ and $x_{2N+1}$ correspond to the left edge of $\mathrm{FM}_1$ and the right edge of $\mathrm{FM}_{N+1}$ respectively), $x_j^-$ is infinitesimally close to $x_j$ on the left, $\hat{Q}_j = \nu(\hat{m}_{j+1} - \hat{m}_j)$ and $\hat{S}_j = \nu(\hat{n}_j - \hat{n}_{j-1})$ are respectively the charge (spin) of the edge segment under $\mathrm{SC}_j$ ($\mathrm{FM}_j$), and $\hat{\varphi}$ is the proximitizing SC order parameter phase. For the first and last FM domains the expression for the hosted spin is different:

$$\hat{S}_1 = \left(\nu\hat{n}_1 - \frac{\nu\hat{\varphi}}{2\pi}\right) - \frac{\sqrt{\nu}}{2\pi}\hat{\theta}_+(x_0), \quad \text{(S19)}$$

$$\hat{S}_{N+1} = \frac{\sqrt{\nu}}{2\pi}\hat{\theta}_+(x_{2N+1}) - \left(\nu\hat{n}_N - \frac{\nu\hat{\varphi}}{2\pi}\right). \quad \text{(S20)}$$

The commutation relations obeyed by the operators can be deduced from the commutation relations for $\hat{\phi}_\uparrow(x)$, $\hat{\phi}_\downarrow(x)$, and their derivatives; the non-trivial commutations are:

$$[\hat{m}_j, \hat{n}_l] = \begin{cases} \frac{i}{\pi\nu} & , j > l, \\ 0 & , j \leq l, \end{cases} \quad \text{(S21)}$$

$$[\hat{Q}_j, \hat{n}_l] = [\hat{S}_j, \hat{m}_l] = \frac{i}{\pi}\delta_{jl}, \quad \text{(S22)}$$

$$[\hat{Q}_j, \hat{S}_l] = \frac{i\nu}{\pi}(\delta_{jl} - \delta_{j+1,l}), \quad \text{(S23)}$$

$$[\hat{Q}_0, \hat{n}_j] = -\frac{i}{\pi}, \; j, l \geq 1. \quad \text{(S24)}$$

The system Hamiltonian then acquires the form

$$H = \sum_{\mathrm{FM}}\left(\frac{\pi v}{2\nu L_{\mathrm{FM}}}\hat{S}_j^2 - tL_{\mathrm{FM}}\cos(2\pi\hat{m}_j)\right)$$
$$+ \sum_{\mathrm{SC}}\left(\frac{\pi v}{2\nu L_{\mathrm{SC}}}\hat{Q}_j^2 - \Delta L_{\mathrm{SC}}\cos(2\pi\hat{n}_j)\right)$$
$$+ \frac{(\hat{Q}_0 - q_g)^2}{2C_{\mathrm{SC}}} + \text{semi-infinite edges}. \quad \text{(S25)}$$

A system described by the first two terms on the r.h.s. of (S25) has been considered in Ref. [3], and the semi-infinite edges in conjunction with such a system have been considered in Refs. [4, 5]. The bulk SC charging energy, which appears in the third term of (S25), is a key ingredient giving rise to the box charging energy, see Eq. (9) of the main text. To the best of our knowledge, it has not been derived in previous works. We note, however, that several recent works [6, 7] have invoked a charging energy as in Eq. (9) of the main text.

An algorithm for finding the low-energy theory for Hamiltonians with large cosine terms as in Eq. (S25) has been put forward in Ref. [8] and we employ it for the analysis presented in Sec. III.3. However, before addressing the problem in its full complexity, it is instructive to consider two toy problems, which we do below.

### III.1. Toy problem 1

Consider a quantum-mechanical problem with

$$H_1 = \frac{\hat{p}^2}{2m} + \frac{K\hat{x}^2}{2} - U_p\cos(2\pi\hat{p}) - U_x\cos(2\pi\hat{x}), \quad \text{(S26)}$$

$$[\hat{p}, \hat{x}] = \frac{i}{\pi\nu} \quad \text{(S27)}$$

with $\nu^{-1}$ being integer. $H_1$ is an illustrative simplification of the first two rows in (S25). First, we solve it ingoring the two middle terms.

$$H_0 = \frac{\hat{p}^2}{2m} - U_x\cos(2\pi\hat{x}). \quad \text{(S28)}$$

Due to Bloch's theorem, the eigenfunctions of $H_0$ satisfy

$$e^{i\pi\nu\hat{p}}\psi_{kn}(x) = \psi_{kn}(x-1) = e^{-ik}\psi_{kn}(x), \quad \text{(S29)}$$

where $k \in [0, 2\pi)$ is the quasi-momentum, $n \in \mathbb{Z}_+$ is the band index, and the eigenfunctions can be written as $\psi_{kn}(x) = e^{ikx}u_{kn}(x)$ with $u_{kn}(x+1) = u_{kn}(x)$. For $U_x \to +\infty$ the potential can be considered as a set of independent cosine minima with weak tunneling between them. Expanding the cosine near its minima and calculating the tunneling between them semi-classically, one finds $u_{kn}(x) = u_n(x) + \delta u_{kn}(x)$, $E_{kn} = E_n + \delta E_{kn}$ with $E_n \approx 2\nu^{-1}\sqrt{U_x/m}(n+1/2) - U_x$ and

$$\delta u_{kn}(x), \delta E_{kn} = O\left(e^{-4\nu\sqrt{mU_x}}\right). \quad \text{(S30)}$$

Since the term $U_p\cos(2\pi\hat{p})$ commutes with $H_0$, it only modifies the eigenenergies $E_{kn} \to E_{kn} - U_p\cos(2\nu^{-1}k)$.

Now we look for eigenfunctions of (S26) in the form

$$\Psi_n(x) = \int_0^{2\pi} dk\, A(k)\psi_{k \mod 2\pi, n}(x). \quad \text{(S31)}$$

Ignoring $O\left(e^{-4\nu\sqrt{mU_x}}\right)$ terms, we obtain

$$\left(-\frac{K}{2}\partial_k^2 - U_p\cos(2\nu^{-1}k) + E_n\right)A(k) = EA(k), \quad \text{(S32)}$$

supplemented by the boundary condition $A(k+2\pi) = A(k)$. In the limit $U_p \to +\infty$, employing the same technique that was used to tackle $H_0$, we find that the eigenfunctions $A_{yl}(k) = e^{-iky}w_{yl}(k)$ have energies $E_{yln} = E_n + \tilde{E}_l + \delta\tilde{E}_{yl}$. Here $y = 0, 1, ..., 2\nu^{-1} - 1$ is the quasi-coordinate, $l \in \mathbb{Z}_+$ is the band index, $w_{yl}(k+\pi\nu) = w_{yl}(k) + \delta w_{yl}(k)$, $\tilde{E}_l \approx 2\nu^{-1}\sqrt{KU_p}(l+1/2) - U_p$, and

$$\delta w_{yl}(x), \delta \tilde{E}_{yl} = O\left(e^{-4\nu\sqrt{U_p/K}}\right). \quad \text{(S33)}$$

Ignoring exponentially small terms, the eigenfunctions satisfy

$$\Psi_{yln}(x) = \int_0^{2\pi} dk\, e^{ik(x-y)}w_l(k)u_n(x), \quad \text{(S34)}$$

$$e^{i\pi\nu\hat{x}}\Psi_{yln}(x) = e^{i\pi\nu y}\Psi_{yln}(x), \quad \text{(S35)}$$

$$e^{i\pi\nu\hat{p}}\Psi_{yln}(x) = \Psi_{y+1 \mod 2\nu^{-1}, ln}(x), \quad \text{(S36)}$$

$$H_1\Psi_{yln}(x) = \left(E_n + \tilde{E}_l\right)\Psi_{yln}(x). \quad \text{(S37)}$$

In particular, the lowest-energy subspace spanned by $\Psi_{y00}(x)$ is $2\nu^{-1}$-fold degenerate up to corrections $O\left(e^{-4\nu\sqrt{U_p/K}}, e^{-4\nu\sqrt{mU_x}}\right)$. Alternatively, one can consider the eigenbasis of $e^{i\pi\nu\hat{p}}$:

$$\tilde{\Psi}_{r00}(x) = \left(2\nu^{-1}\right)^{-1/2}\sum_y e^{-i\pi\nu ry}\Psi_{y00}(x),$$

$$r = 0, 1, ..., 2\nu^{-1} - 1. \quad \text{(S38)}$$

### III.2. Toy problem 2

Consider the Hamiltonian

$$H_2 = H_1 + \frac{\hat{p}_0^2}{2m_0}, \quad \text{(S39)}$$

where $H_1$ is defined in (S26), and $\hat{p}_0$ satisfies

$$[\hat{p}_0, \hat{x}] = -\frac{i}{\pi\nu}, \quad [\hat{p}_0, \hat{p}] = 0. \quad \text{(S40)}$$

We note that $H_1$ consists of a simplification of the first two r.h.s. terms in (S25), while $\hat{p}_0^2/2m_0$ is a simplified analogue of the bulk SC charging energy term. In analogy with the SC charge, $\hat{Q}_0$ whose conjugate operator is $\hat{\varphi}$, for $\hat{p}_0$ there is a conjugate variable $\hat{x}_0$: $[\hat{p}_0, \hat{x}_0] = i$. For simplicity, in the present example we consider the operator $\hat{p}_0$ which has a continuous unbounded spectrum. This is to be contrasted with $\hat{Q}_0$, whose spectrum is discrete.

Due to the presence of the second term in (S39), one cannot directly use the solution of $H_1$ from the previous section. Defining

$$\hat{p}_+ = \hat{p} + \hat{p}_0, \quad [\hat{p}_+, \hat{x}] = 0, \quad \text{(S41)}$$

one can rewrite

$$\frac{\hat{p}^2}{2m} + \frac{\hat{p}_0^2}{2m_0} = \frac{\hat{p}_+^2}{2m_+} + \frac{(\hat{p} - \alpha\hat{p}_+)^2}{2m_p}, \quad \text{(S42)}$$

$$m_+ = m + m_0, \quad \alpha = \frac{m}{m+m_0}, \quad m_p = \alpha m_0. \quad \text{(S43)}$$

Therefore,

$$H_2 = \tilde{H}_1 + \frac{\hat{p}_+^2}{2m_+}, \tag{S44}$$

$$\tilde{H}_1 = \frac{(\hat{p} - \alpha\hat{p}_+)^2}{2m_p} + \frac{K\hat{x}^2}{2} - U_p\cos(2\pi\hat{p}) - U_x\cos(2\pi\hat{x}). \tag{S45}$$

There are two differences between $\tilde{H}_1$ and $H_1$: first, the renormalized $m_p$ instead of $m$, and second, the $\alpha\hat{p}_+$ shift. Since $\left[\tilde{H}_1, \hat{p}_+\right] = 0$ both operators can be diagonalized simultaneously. For every eigenstate of $\hat{p}_+$, with a given eigenvalue, one can diagonalize $\tilde{H}_1$ following the procedure of Sec. III.1. The $\alpha\hat{p}_+$ shift in Eq. (S45) leads only to exponentially small corrections. Therefore, the low-energy Hilbert space of the system is spanned by $|p_+, y\rangle$ with $y$ labeling the exponentially degenerate ground states of $\tilde{H}_1$, and the low-energy effective Hamiltonian is

$$H_{\text{eff}} = \frac{\hat{p}_+^2}{2m_+}, \tag{S46}$$

where we have omitted the constant offset by the ground-state energy of $\tilde{H}_1$.

### III.3. Solving (S25) in the limit $t, \Delta \to +\infty$

A general algorithm for solving Hamiltonians with large cosine terms as in Eq. (S25) has been developed in Ref. [8]. According to this algorithm, the cosine arguments $\{\hat{C}_q\} = \{2\pi\hat{m}_j, 2\pi\hat{n}_j\}$ play the key role. The low-energy Hilbert space is obtained through restricting the original problem Hilbert space by demanding $\cos\hat{C}_q|\psi\rangle = |\psi\rangle$, which implies that the allowed low-energy operators are those and only those that commute with all $\cos\hat{C}_q$. The low-energy effective Hamiltonian then assumes the form

$$H_{\text{eff}} = H_0 - \frac{1}{2}\sum_{q,r}\left(\mathcal{M}^{-1}\right)_{qr}\hat{\Pi}_q\hat{\Pi}_r, \tag{S47}$$

where $H_0$ is the quadratic part of the original Hamiltonian, and the variables $\hat{\Pi}_q$, conjugate to $\hat{C}_q$, are

$$\hat{\Pi}_q = \frac{1}{2\pi i}\sum_r \mathcal{M}_{qr}\left[\hat{C}_r, H_0\right], \tag{S48}$$

where $\mathcal{M}_{qr}$ is defined through

$$\mathcal{M} = \mathcal{N}^{-1}; \quad \mathcal{N}_{qr} = -\frac{1}{4\pi^2}\left[\hat{C}_q, \left[\hat{C}_r, H_0\right]\right]. \tag{S49}$$

Finally, the non-trivial commutation relations $\left[\hat{C}_q, \hat{C}_r\right]$ lead to a degenerate low-energy Hilbert space, while the finite coefficients in front of the cosine terms lead to splitting between these degenerate states, as also discussed in Ref. [8].

Applying this algorithm to the two toy problems discussed in Secs. III.1, III.2, one recovers the results obtained there. We now discuss the application of this algorithm to Hamiltonian (S25). The admissible low energy operators, i.e., those that commute with all $\cos 2\pi\hat{m}_j$ and $\cos 2\pi\hat{n}_j$, are $e^{i\pi\nu\hat{m}_j}$, $e^{i\pi\nu\hat{n}_j}$, $\partial_x\hat{\phi}_{\uparrow/\downarrow}$ and $e^{i\sqrt{\nu}\hat{\phi}_{\uparrow/\downarrow}}$ operators at the semi-infinite edges, and the total box charge $\hat{Q}_{\text{tot}} = \hat{Q}_0 + \sum_{j=1}^N \hat{Q}_j$. All the other low-energy operators can be built out of these. The effective low-energy Hamiltonian is

$$H_{\text{eff}} = \frac{1}{2C_{\text{box}}}\left(\hat{Q}_{\text{tot}} - q_g\right)^2 + \sum_{s=\pm 1}\left(\int_{-\infty}^{x_0} + \int_{x_{2N+1}}^{+\infty}\right)dx(\partial_x\hat{\phi}_s)^2, \tag{S50}$$

where $C_{\text{box}} = C_{\text{SC}} + N\nu L_{\text{SC}}/(\pi v)$. The second term in $H_{\text{eff}}$ represents the Hamiltonian of the semi-infinite edges that are glued together by the conditions

$$\hat{\phi}_\uparrow(x_0) - \hat{\phi}_\downarrow(x_0) = 2\pi\sqrt{\nu}\hat{m}_1, \tag{S51}$$

$$\hat{\phi}_\uparrow(x_{2N+1}) - \hat{\phi}_\downarrow(x_{2N+1}) = 2\pi\sqrt{\nu}\hat{m}_{N+1}. \tag{S52}$$

Therefore, the fields at the edges can be redefined so that the part to the left of $\text{FM}_1$ is described as a free FQH edge, and the part to the right of $\text{FM}_{N+1}$ is described as another one.

The degeneracy due to the non-commutation of $\hat{m}_j$, $\hat{n}_l$ and the expression of parafermionic operators through $\hat{m}_j$, $\hat{n}_l$ is described the same way as without the charging energy [3, 4]. The box states can be labeled by eigenvalues of operators $e^{i\pi\hat{Q}_j}$ and $\hat{Q}_{\text{tot}}$. However, now not all ($\hat{Q}_j \mod 2$) are independent as $\hat{Q}_0$ is an integer multiple of 2, thus $\prod_j e^{i\pi\hat{Q}_j} = e^{i\pi\hat{Q}_{\text{tot}}}$. Since the variables of the semi-infinite edges commute with $\hat{Q}_j$ and $\hat{Q}_{\text{tot}}$, the edges decouple from the box. Therefore, the box can be described independently of the edges by the Hamiltonian in Eq. (9) of the main text. The lifting of the degeneracy is $O\left(\exp\left[-4L_{\text{SC}}(\frac{\nu\Delta}{\pi v})^{1/2}\left(\frac{\pi v C_{\text{SC}}}{\pi v C_{\text{SC}} + \nu L_{\text{SC}}}\right)^{1/2}\right]\right)$, $O\left(\exp\left[-4L_{\text{FM}}(\frac{\nu t}{\pi v})^{1/2}\right]\right)$.

Finally, we would like to note that similar results can be derived for a generic capacitive coupling between the box elements:

$$H_C = \frac{(\hat{Q}_0 - q_g)^2}{2C_{\text{SC}}} + \sum_{jl}\left[A_{jl}\hat{Q}_j\hat{Q}_l + B_{jl}\hat{S}_j\hat{S}_l + C_{jl}\hat{Q}_j\hat{S}_l\right] + \hat{Q}_0\sum_j\left[d_j\hat{Q}_j + e_j\hat{S}_j\right], \tag{S53}$$

with arbitrary coefficients $A_{jl}, B_{jl}, C_{jl}, d_j, e_j$ (provided that the quadratic in charges and spins part of $H_C$ is positive definite). In this case the low-energy expression for the box Hamiltonian, Eq. (9) of the main text, retains the same form. However, the precise expression for the box capacity $C_{\text{box}}$ changes in this case, and the gate parameter $q_g$ should be replaced by $q_{\text{box}} = \text{const} \times q_g$.

## IV. SOME DETAILS REGARDING THE TWO-LEAD MEASUREMENT SCHEME

Here we discuss how the two-lead measurement scheme discussed in the main text implements a projective measurement. We assume that $H_{\text{T}12}$, see Eq. (11) of the main text, is switched on at time $t = 0$. Suppose the initial state of the $N$-box is $|\psi\rangle = \sum_{r,\{X\}} A_{r,\{X\}}|r,\{X\}\rangle$, where $|r, \{X\}\rangle$ is a basis of the box low-energy states such that

$$\hat{\alpha}_{j_2,s}\hat{\alpha}^\dagger_{j_1,s}|r,\{X\}\rangle = e^{i\nu(\pi r + \delta\varphi/2)}|r,\{X\}\rangle, \quad \text{(S54)}$$

and $\{X\}$ denotes all other quantum numbers labeling the box state. Below we calculate the box reduced density matrix $\hat{\rho}_{\text{box}}(t)$ as a function of time. We find that

$$\langle r,\{X\}|\hat{\rho}_{\text{box}}(t)|r',\{X'\}\rangle = A^*_{r',\{X'\}}A_{r,\{X\}}e^{-F_{rr'}(t)}, \quad \text{(S55)}$$

$$\text{Re}\, F_{rr'}(t) = t|\eta_r - \eta_{r'}|^2 v^{-2\nu}\frac{\pi(\nu|V_{12}|)^{2\nu-1}}{\Gamma(2\nu)}, \quad \text{(S56)}$$

where $\eta_r = \eta_{\text{ref}} + \eta_{\text{cot}}e^{i\nu(\pi r + \delta\varphi/2)}$, $\Gamma(x)$ is the Euler gamma function, and the last expression is valid for $t \gg 1/(\nu|V_{12}|)$. Therefore, the scheme kills the density matrix elements that are off-diagonal in $r$ while preserving coherent superposition in the rest of quantum numbers, thus implementing a projective measurement of $r$.

The derivation of the above result is as follows. Evolution of the leads and box state can be described in the interaction representation by evolution operator

$$\hat{U}(t) = T\exp\left(-i\int_0^t dt H_{\text{T}12}\right). \quad \text{(S57)}$$

The initial density matrix of the system is $\hat{\rho}(t=0) = \hat{\rho}_{\text{box}}(0) \otimes \hat{\rho}_{\text{leads}}(0)$ with $\hat{\rho}_{\text{box}}(0) = |\psi\rangle\langle\psi|$ and $\hat{\rho}_{\text{leads}}(0) \sim \exp\left(-\sum_{i=1,2} H_i/T\right)$, where $H_i$ are the leads' Hamiltonians and $T$ is the system temperature. Therefore, the reduced density matrix of the box at time $t$ is

$$\hat{\rho}_{\text{box}}(t) = \text{Tr}_{\text{leads}}\left[\hat{U}(t)\hat{\rho}(0)\hat{U}^\dagger(t)\right]$$
$$= \sum_{r,r'}\sum_{\{X\},\{X'\}} |r,\{X\}\rangle\langle r',\{X'\}|$$
$$\times A^*_{r',\{X'\}}A_{r,\{X\}}\langle U^\dagger_{r'}(t)U_r(t)\rangle_{\text{leads}}, \quad \text{(S58)}$$

where $U_r(t)$ is obtained from $\hat{U}(t)$ by substituting $\hat{\alpha}_{j_2,s}\hat{\alpha}^\dagger_{j_1,s}$ in $H_{\text{T}12}$ with its eigenvalue (S54). We define

$$F_{rr'}(t) = -\ln\langle U^\dagger_{r'}(t)U_r(t)\rangle_{\text{leads}}. \quad \text{(S59)}$$

Since $U^\dagger_r(t)U_r(t) = 1$, $F_{rr}(t) = 0$. For the off-diagonal components we perform a second order cumulant expansion in $H_{\text{T}12}$ and obtain

$$\text{Re}\, F_{rr'}(t) = 2|\eta_r - \eta_{r'}|^2 v^{-2\nu}\cos\pi\nu\times$$
$$\int_0^t d\tau(t-\tau)\cos(\nu V_{12}\tau)\left(\frac{\pi T}{\sinh \pi T\tau}\right)^{2\nu}. \quad \text{(S60)}$$

The last expression is valid for $\nu < 1/2$, it should be analytically continued in order to obtain the answer for $\nu \geq 1/2$. For $|V_{12}| \gg T$ one obtains

$$2\cos\pi\nu\int_0^t d\tau(t-\tau)\cos(\nu V_{12}\tau)\left(\frac{\pi T}{\sinh \pi T\tau}\right)^{2\nu}$$
$$= 2\cos\pi\nu\int_0^t d\tau(t-\tau)\cos(\nu V_{12}\tau)\tau^{-2\nu}$$
$$= \frac{\pi(\nu|V_{12}|)^{2\nu-1}}{\Gamma(2\nu)}t\times\left(1 + O\left([\nu|V_{12}|t]^{-\min(2\nu,1)}\right)\right), \quad \text{(S61)}$$

which leads to Eq. (S56). As an analytic function, this answer is valid for any $\nu > 0$, including $\nu = 1$. For $\nu = 1$ our results reproduce those of Ref. [9], with $v^{-\nu}$ playing the role of the density of states of a lead.

It is interesting to note that the dephasing rate (S56), and more generally (S60), does not depend on the value of $\eta_{\text{ref}}$. It is a consequence of the fact that each quasiparticle that passes through the box causes a state-dependent phase rotation of the box state, and those that tunnel through the reference arm do not. Dephasing is generated by the inability to track the number of quasiparticles that have passed through the box. In the case of $\eta_{\text{ref}} = 0$, one is not able to distinguish different box states by measuring $I_{\text{T}12}$, as can be seen from Eqs. (11,12) of the main text. Yet we note that dephasing does take place, albeit one can restore coherence between different box states by counting the number of quasiparticles transferred between the leads. For $\eta_{\text{ref}} \neq 0$ one cannot restore phase coherence by simply counting the number of quasiparticles as each transferred quasiparticle could have traveled either through the box or through the reference arm. The relation of the dephasing to the number of quasiparticles transferred through the box results in the same voltage dependence of the dephasing rate (S56) and the current $I_{\text{T}12}$ (12).

Having in mind possible experimental realizations, we estimate the typical decay time of the off-diagonal matrix element for $\nu = 1/3$ to be $\approx 10\,\text{ns}$ when $\eta_{\text{ref}} \approx \eta_{\text{cot}}$ and $I_{\text{T}12} \approx 10\,\text{pA}$, which is a realistic value of tunneling current. Under these conditions, the measurement time

is limited by the time required to accumulate sufficient statistics in order to distinguish the values of $I_{\text{T}12}$ for different $r$.

## V. THE ALGEBRA OF OPERATORS ACTING IN THE 2-BOX DEGENERATE SUBSPACE AND THEIR EIGENSTATES

The 2-box degenerate subspace is spanned by states $|Q_1 \mod 2\rangle$ that are eigenstates of operator $e^{i\pi \hat{Q}_1}$. All the operators of the form $\hat{\alpha}_{j_2,s}\hat{\alpha}^\dagger_{j_1,s}$ acting in this subspace can be expressed through $e^{i\pi \hat{Q}_1}$ and $e^{i\pi \hat{S}_2}$. It is of importance, therefore, to study the properties of these two operators.

It follows from Eq. (S23) that

$$\left[\hat{Q}_1, e^{i\pi \hat{S}_2}\right] = \nu e^{i\pi \hat{S}_2}. \quad (S62)$$

Using this fact and the arbitrariness of choosing the phases of the basis states $|Q_1 \mod 2\rangle$, one can define

$$|Q_1 \mod 2\rangle = e^{i\pi\nu^{-1}(Q_1 \mod 2)\hat{S}_2}|Q_1 \mod 2 = 0\rangle, \quad (S63)$$

$$e^{i\pi \hat{S}_2}|Q_1 \mod 2\rangle = |Q_1 + \nu \mod 2\rangle. \quad (S64)$$

The eigenstates $|S_2 \mod 2\rangle$ of $e^{i\pi \hat{S}_2}$ have the form

$$|S_2 \mod 2\rangle = \frac{1}{\sqrt{n}} \sum_{r=0}^{n-1} e^{-i\pi r(S_2 \mod 2)} \times |Q_1 \mod 2 = \nu r\rangle \quad (S65)$$

with $n = 2/\nu$. Conversely, the $|Q_1 \mod 2\rangle$ states can be expressed through $|S_2 \mod 2\rangle$ via relation (13) in the main text.

## VI. $N$-BOX STATE MANIPULATION WITH QADS

Here we elaborate on how transferring a quasiparticle between two QADs each of which is connected to one parafermion allows one to manipulate the state of an $N$-box. Such a manipulation involves two QADs, each described by Hamiltonian (14). Due to coupling to the parafermions, a fractional quasiparticle can hop between the QADs through the box, which is described by the effective cotunneling Hamiltonian $H_{\text{cot,QAD}}$. Consider the initial state of the QADs being $|1\rangle_1|0\rangle_2$, i.e., QAD1 is filled and QAD2 is empty. The box states can be expanded in $|r, \{X\}\rangle$, the eigenbasis of $\hat{\alpha}_{j_2,s}\hat{\alpha}^\dagger_{j_1,s}$ (S54), where $\hat{\alpha}_{j_1,s}$ and $\hat{\alpha}_{j_2,s}$ are the parafermions to which QAD1 and QAD2 are coupled respectively.

The cotunneling Hamiltonian transforms

$$|1\rangle_1|0\rangle_2|r, \{X\}\rangle \rightleftarrows |0\rangle_1|1\rangle_2|r, \{X\}\rangle. \quad (S66)$$

Therefore, the problem splits into a set of $2 \times 2$ problems each described by

$$H_r = \begin{pmatrix} \Delta V/n & \eta^*_{\text{cot}} e^{-i\nu(\pi r + \delta\varphi/2)} \\ \eta_{\text{cot}} e^{i\nu(\pi r + \delta\varphi/2)} & -\Delta V/n \end{pmatrix}, \quad (S67)$$

where $\Delta V = V_{\text{QAD1}} - V_{\text{QAD2}}$.

Consider the Landau-Zener problem $\Delta V = \lambda t/\nu$ for this Hamiltonian. Then the initial state $|\psi(t_i \to -\infty)\rangle = |1\rangle_1|0\rangle_2|r, \{X\}\rangle$ at large times becomes

$$|\psi(t \to +\infty)\rangle = e^{i\varphi_0(t_i)} \bigg[ e^{-\pi\gamma - i\varphi_0(t)}|1\rangle_1|0\rangle_2|r, \{X\}\rangle$$
$$+ \frac{\sqrt{2\pi\gamma}}{\Gamma(1+i\gamma)} e^{-\frac{1}{2}\pi\gamma + i\varphi_0(t)} \frac{\eta_{\text{cot}}}{|\eta_{\text{cot}}|} e^{i\nu(\pi r + \delta\varphi/2)}|0\rangle_1|1\rangle_2|r, \{X\}\rangle \bigg], \quad (S68)$$

where $\gamma = |\eta_{\text{cot}}|^2/\lambda$, and

$$\varphi_0(t) = \frac{\lambda t^2}{4} + \frac{1}{2}\gamma \ln(\lambda t^2) - \frac{3\pi}{8}. \quad (S69)$$

In particular, the probability of not transferring a quasiparticle between the QADs is

$$P_{\text{LZ}} = \exp(-2\pi\gamma). \quad (S70)$$

Therefore, by changing the QAD voltages slowly ($\lambda \to 0$, $\gamma \to +\infty$), one can transfer the quasiparticle with probability 1; the system final state will then be

$$e^{i\nu(\pi r + \delta\varphi/2)}|0\rangle_1|1\rangle_2|r, \{X\}\rangle = |0\rangle_1|1\rangle_2\hat{\alpha}_{j_2,s}\hat{\alpha}^\dagger_{j_1,s}|r, \{X\}\rangle, \quad (S71)$$

up to an $r$-independent overall phase. Moreover, if one performs the operation with non-vanishing speed $\lambda$ and measures the QADs' state in the end one will find the system in one of the states

$$|1\rangle_1|0\rangle_2|r, \{X\}\rangle \text{ or } |0\rangle_1|1\rangle_2\hat{\alpha}_{j_2,s}\hat{\alpha}^\dagger_{j_1,s}|r, \{X\}\rangle \quad (S72)$$

up to $r$-independent overall phases with probabilities $P_{\text{LZ}}$ and $1 - P_{\text{LZ}}$ respectively. One can repeat the operation until the transfer of a quasiparticle has happened. One way or the other, transferring a quasiparticle between the dots implements the action of the $\hat{\alpha}_{j_2,s}\hat{\alpha}^\dagger_{j_1,s}$ operator on the $N$-box low-energy Hilbert space. Importantly, this operation is applied independently of microscopic details of the process.

## VII. THE APPLICABILITY OF OUR RESULTS TO $\nu = 2/3$ PARAFERMIONS

The results stated in the main text are applicable, with minor corrections, also to parafermions implemented through $\nu = 2/3$ FQH edges. Here we discuss this issue in some detail. A scheme to implement parafermions in a system involving a pair of spin-unpolarized $\nu = 2/3$

FQH puddles has been proposed in Ref. [10]. The setup there is similar to the setup for obtaining parafermions at $\nu = 1/(2k+1)$: it involves two FQH edges that are gapped in different regions by either electron tunneling or SC pairing between the edges. The domain walls between these regions give rise to parafermionic operators.

The $\nu = 2/3$ edge has a more complicated structure than a $\nu = 1/(2k+1)$ edge. Namely, it has a charged mode $\hat{\phi}_c$ that is described similarly to the only mode present for $\nu = 1/(2k+1)$ and also a neutral mode $\hat{\phi}_n$ [11, 12] (see, however, Refs. [13, 14]). In the scheme of Ref. [10], the neutral mode is uniformly gapped in the whole system and at low energies only the charged mode is of importance. Therefore, the parafermion operators are low energy projections of the elementary excitation $e^{i\sqrt{\nu}\hat{\phi}_c}$ that carries the electric charge $\nu e$ and does not excite the neutral mode. Thus, PFs for $\nu = 2/3$ can be described with the same formalism as for $\nu = 1/(2k+1)$.

QADs in $\nu = 2/3$ can host both purely charged $2e/3$ excitations and excitations that have $e/3$-charge and also excite the neutral mode. Since the latter ones cannot travel through the system of parafermions (the neutral mode is gapped), our results regarding the QAD manipulation of parafermions are directly applicable to the case of $\nu = 2/3$.

The results for measuring the state of parafermions with leads are also applicable, up to a minor modification. The measurement scheme we propose involves not just cotunneling through the PF box but also direct tunneling between the leads (involving the reference arm), in which both the purely charged excitations and those having a neutral component participate. Therefore, the expression for the tunneling current in Eq. (12) of the main text should be modified by replacing $|\hat{\eta}_{T12}|^2 \to |\hat{\eta}_{T12}|^2 + |\eta_n|^2$, where $|\eta_n|^2$ characterizes the strength of tunneling of all the neutral-carrying excitations. Thus, for a given voltage $V_{12}$, the current $I_{T12}$ gets offset by a constant current, still enabling one to distinguish different states of parafermions. The results for dephasing of the box's reduced density matrix (S55), (S56), are applicable without changes, which is easy to understand: neutral-carrying excitations cannot pass through the system of parafermions and therefore do not contribute to dephasing.

---



\end{document}